\documentclass[pdflatex,sn-apa]{sn-jnl}

\usepackage{graphicx}%
\usepackage{multirow}%
\usepackage{amsmath,amssymb,amsfonts}%
\usepackage{amsthm}%
\usepackage{mathrsfs}%
\usepackage[title]{appendix}%
\usepackage{xcolor}%
\usepackage{textcomp}%
\usepackage{manyfoot}%
\usepackage{booktabs}%
\usepackage{algorithm}%
\usepackage{algorithmicx}%
\usepackage{algpseudocode}%
\usepackage{listings}%
\usepackage{soul}%
\usepackage{array}
\usepackage{makecell}
\usepackage{booktabs}

\begin{document}

\title[Article Title]{Time-resolved X-ray radiography of through-thickness liquid transport in partly saturated needle-punched nonwovens}

\author[1,2]{\fnm{Patrick} \sur{Wegele}}

\author[3]{\fnm{Zisheng} \sur{Yao}}

\author[1]{\fnm{Jonas} \sur{Tejbo}}

\author[3]{\fnm{Julia K.} \sur{Rogalinski}}

\author[1]{\fnm{Tomas} \sur{Rosén}}

\author[4]{\fnm{Alexander} \sur{Groetsch}}

\author[5]{\fnm{Kim} \sur{Nyg\aa rd}}

\author[5]{\fnm{Eleni Myrto} \sur{Asimakopoulou}}

\author[3]{\fnm{Pablo} \sur{Villanueva-Perez}}

\author*[1]{\fnm{L. Daniel} \sur{Söderberg}}\email{dansod@kth.se}

\affil[1]{\orgdiv{Department of Fibre and Polymer Technology}, \orgname{KTH Royal Institute of Technology}, \orgaddress{\street{Teknikringen 56}, \city{Stockholm}, \postcode{SE-10044}, \country{Sweden}}}

\affil[2]{\orgname{J.M. Voith SE \& Co. KG}, \orgaddress{\street{St. Poeltener Strasse 43}, \city{Heidenheim an der Brenz}, \postcode{89522}, \country{Germany}}}

\affil[3]{\orgdiv{Synchrotron Radiation Research and NanoLund}, \orgname{Lund University}, \city{Lund}, \country{Sweden}}

\affil[4]{\orgdiv{Department of Engineering Mechanics}, \orgname{KTH Royal Institute of Technology}, \city{Stockholm}, \country{Sweden}}

\affil[5]{\orgdiv{MAX IV Laboratory}, \orgname{Lund University}, \city{Lund}, \country{Sweden}}

\abstract{
Nonwoven fibre networks underpin filtration, insulation and geotextiles, where liquid uptake, redistribution and release govern performance. In needle-punched felts, barbed needles mechanically entangle fibres and partially reorient them toward the thickness direction~($z$), creating out-of-plane “pillars” and heterogeneity. While mechanical and structural consequences of needling are well documented, dynamic $z$-direction transport in partly saturated networks remains difficult to access due to opacity and sub-second timescales. Here we combine micro-CT~(µCT) of dry structure with time-resolved X-ray radiography during droplet addition to quantify through-thickness transport as a function of saturation and needling intensity, using a compact Washburn-type descriptor for dynamics. Results show an exponential dependence of $z$-directional liquid transport on saturation, consistent with previous models for in-plane relative permeability of nonwoven networks. Additionally, increased needle-punch intensity reorients fibres toward the $z$-direction, forming preferential flow pathways that enhance through-thickness transport, even as single-phase permeability decreases. These findings underscore needle-punch as a key design parameter for tuning liquid transport in nonwoven fibre networks. The approach provides an experimental and modelling framework for dynamic, capillarity-driven transport in opaque fibrous materials.

}




\maketitle

\section{Introduction}\label{sec:Sec_Introduction}
Nonwoven fibre networks are widely used in applications requiring liquid handling and transport, such as filtration, insulation and geotextiles. To meet the desired product properties in the individual application, detailed knowledge about the interaction of the network with the fluid is indispensable. This is a complicated task as both the network's internal structure and the relative network saturation determine the overall network permeability. Many nonwoven networks are produced in a needle-punch process, where barbed needles penetrate the fibre web, mechanically entangling the fibres and partially reorienting them from the in-plane to the out-of-plane direction, which we call by definition the $z$-direction. This technique alters the internal structure by creating fibre "pillars" in the $z$-direction along needle tracks. While essential for structural integrity, the effects of needle-punch on the liquid transport properties of the resulting networks remain underexplored.

Previous studies have primarily focused on the influence on the mechanical properties of needle-punched networks \citep{Miao2004b, Jearanaisilawong2008, Martinez2015, Asri2023} or structural changes due to needling \citep{Ishikawa2019, Yu2017, Miao2004a}. It has been pointed out that increasing the punch-needling intensity causes a higher network compaction, which results in a lower network porosity \citep{Shadin2021}. Additionally, there is less information available regarding the liquid transport properties. \cite{Debnath2020} investigated the effect of needle-punching intensity and depth on the air permeability of nonwoven polyester networks, finding that the permeability decreases with increasing needle-punch intensity. \cite{Asis2009} came to similar results, which were explained by a reduction of the size of big and medium-sized pores by increasing fibre entanglement, which determines the permeability. This aligns with later findings obtained by \cite{Li2016}.

Capillary-driven liquid transport in porous media is traditionally modelled using the Lucas–Washburn equation~\citep{Lucas1918, Washburn1921}, which treats the medium as a bundle of capillaries. Consequently, several extensions of the Lucas-Washburn equation have been formulated to account for the effect of inertial forces~\citep{Bosanquet1923}, network tortuosity \citep{HODGSON1988} and fibre swelling~\citep{Hoyland1977, Schuchardt1991, Masoodi2010}. For unsaturated flow, the Richards equation~\citep{Richards1931} provides an approach that allows for the numerical calculation of liquid imbibition into fibrous networks \citep{ASHARI2010, Zarandi2018}.

A key concept in liquid flow in unsaturated porous media is the relative permeability, which scales the single-phase permeability according to saturation with the wetting phase. In partially saturated systems, the total liquid permeability $K_{ii}$ in a principal direction is commonly expressed as the product of the single-phase permeability $K^s_{ii}$ at full network saturation ($\Theta=1$) and the relative permeability $K^r_{ii}(\Theta)$ ($0<K^r_{ii}(\Theta)<1$), which depends on the network saturation $\Theta$ \citep{Masoodi2012, Dullien2012, ASHARI2010}:

\begin{equation}
    K_{ii}=K^s_{ii}\cdot K^r_{ii}(\Theta)
\end{equation}

While empirical relations exist for granular media \citep{Brooks1965, vanGenuchten1980}, limited data are available for fibrous networks. \cite{Landeryou2005} and \cite{Ashari2009A} found exponential relationships between saturation and in-plane permeability. Directional differences in permeability have also been reported, with in-plane flow often twice as high as $z$-directional flow due to lower resistance along fibres \citep{Ashari2009B}.  Regarding needle-punched nonwoven fibre networks, \cite{Landeryou2005} showed that the relative in-plane permeability follows a cubic exponential relationship with the saturation. This finding was extended by \cite{Ashari2009A} for straight fibre networks. For layered nonwoven fibre networks, it was shown that the in-plane permeability is almost two times higher than the $z$-directional permeability \citep{Ashari2009B}, which is explained by a lower flow resistance of the liquid phase to flow along the fibres then crossing them \citep{ASHARI2010}. While also directional formulations of permeability have been proposed \citep{Mao2003, Mao2009}, their application is limited due to the heterogeneous and anisotropic nature of real networks. Consequently, several experimental approaches are commonly used to determine the liquid transport properties in low-density fibrous media \citep{Abedsoltan2023}.

Although static liquid distribution in needle-punched networks has been studied \citep{Wegele2024}, dynamic $z$-directional liquid transport remains difficult to assess \citep{Landeryou2003} due to the opacity of the material and the timescales these processes happen in. As optical methods \citep{ARORA2006} are ineffective in the thickness direction, X-ray-based techniques are the method of choice. Building on earlier work showing the effectiveness of X-ray imaging \citep{Landeryou2005, Wegele2024} this study aims to answer the following questions:

\begin{itemize}
    \item How does $z$-directional liquid transport in needle-punched nonwoven fibre networks depend on saturation?
    \item How does varying needle-punch intensity influence the $z$-directional liquid transport?
    \item Is it possible to evaluate the data using a Lucas–Washburn-based model, thereby benefiting from the low complexity of the formulation?
\end{itemize}

\section{Materials}
Within our analysis, we are using multilayer fibre networks similar to those described by \cite{Thibault2001}, \cite{Thibault2002} and \cite{Wegele2024}. The networks are formed by several nonwoven polyamide~6~(PA6) fibre layers of varying fibre diameters, needle-punched on a woven PA6 base layer. As indicated in Table~\ref{tab:sampleOverview}, three fibre network types are examined, differing in structural configuration and needle-punch intensity. Sample~\textit{"Ref"} represents an industrially used configuration and serves as a reference for assessing $z$-directional liquid transport as a function of saturation. Samples~\textit{"High-NPI"} and~\textit{"Low-NPI"} have simpler structures and differ only in the needle-punch intensity (\textit{"NPI"}, quantified in penetrations per cm²) applied during manufacturing. The bulk porosity $\phi_B$ of the individual networks is calculated using the grammage $w$, the density of the fibre material $\rho_f$ and the height of the individual network $h$:

\begin{equation}
    \phi_B=1-\frac{w}{\rho_{f}-h}
\end{equation}

\begin{table}[h]
\caption{Sample overview. Data are determined on the macroscopic sheet before cutting out samples for further analysis and testing.}
\label{tab:sampleOverview}
\centering
\begin{tabular}{@{}lccc@{}}
\toprule
 & Ref. & High-NPI & Low-NPI \\ \midrule
Grammage $w$ {[}g/m²{]} & 1480 & 1440 & 1440 \\
Penetrations / {[}1/cm²{]}    & 750 & 1200 & 700 \\
Batt fibre layers    & 2 & 1 & 1 \\
Batt fibre diameter {[}µm{]} &  39; 63 & 35 & 35 \\
Sample height $h$ {[}mm{]} & 2.63 & 2.67 & 2.86 \\
Sample diameter $D$ {[}mm{]} & 10 & 8 & 8 \\
Bulk porosity $\phi_B$  & 0.502 & 0.523 & 0.554 \\
\bottomrule
\end{tabular}
\end{table}

\begin{figure}[htbp]
\centerline{\includegraphics[width=\textwidth]{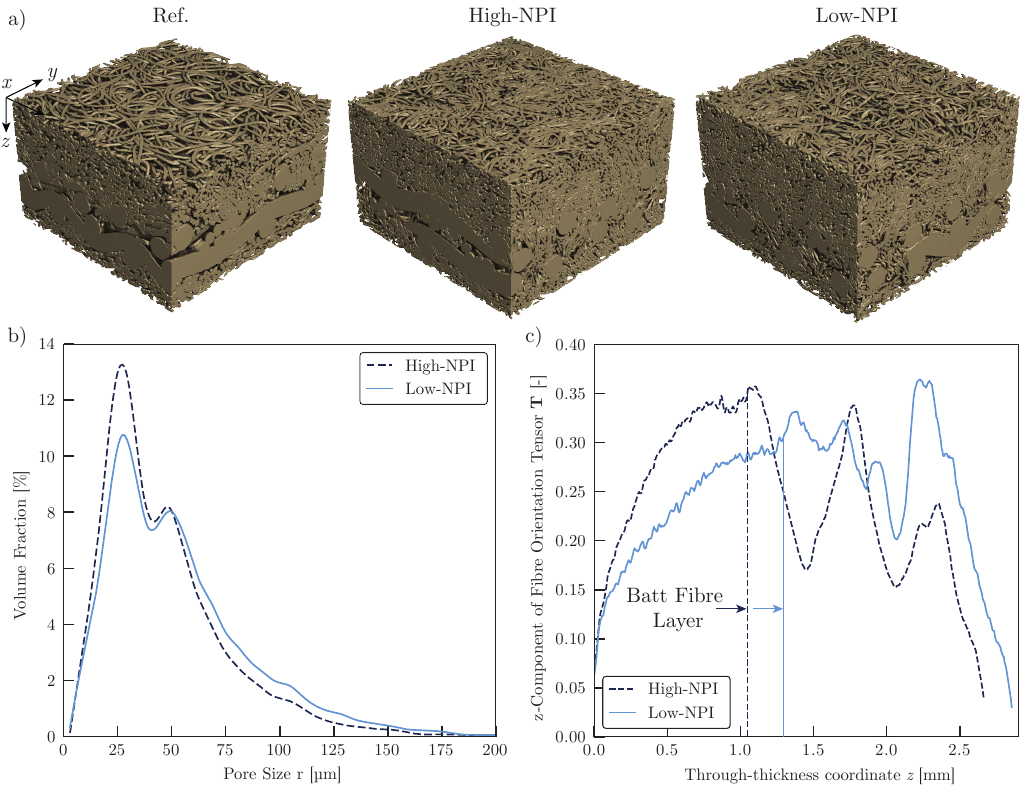}}
\caption{Tomographic reconstruction and analysis of the samples used in the underlying experiment. See Appendix~\ref{Sec:App0} for an uncertainty analysis of the dry network evaluations. a) Rendering of the reconstruction of samples Ref,~High-NPI and~Low-NPI. b) Pore size distribution of samples~High-NPI and~Low-NPI. Sample~Ref is not investigated in terms of pore size and fibre orientation due to the differing composition, which does not allow a comparison to samples~High-NPI and Low-NPI c) $zz$-component of the fibre orientation tensor as a function of the through-thickness coordinate $z$.}
\label{Fig:PoresOrientation}
\end{figure}

All samples are extracted from the networks using a circular punch and are reused in both static and dynamic X-ray experiments to minimise variability. To better understand the network structure, we perform static µCT~measurements at a voxel resolution of 6~µm of the samples~Ref,~High-NPI and~Low-NPI in a Pheonix Nanotom~M nanofocus X-ray µCT~system (GE~Sensing \& Inspection Technologies~GmbH) with previously reported parameters \citep{Wegele2024}. To mitigate beam hardening, a 0.5~mm aluminium filter is used to suppress the low-energy component of the polychromatic cone beam. Based on the tomographic reconstructions, segmentation in fibres and air is done using the  Otsu~method \citep{Otsu1979}. The tomographic reconstructions as well as the pore size distribution and the fibre orientation as $zz$-component of the orientation tensor $T_{ij}(z)$, are visible in Figure~\ref{Fig:PoresOrientation}. The pore size distribution is calculated using a granulometry approach as described by \cite{Silin2006} and implemented by \cite{Schulz2007}, while $T_{zz}(z)$ is calculated based on a Star Length Distribution~(SLD) approach implemented in the commercial software GeoDict \citep{Becker2021}.

Regarding samples~High-NPI~and~Low-NPI it is apparent that increasing the needle-punch intensity reduces the amount of big- and medium-sized pores~($r>$50~µm) while the amount of evaluated small pores~(25~µm$<r<$50~µm) is increased. This aligns with observations for the needle-punch penetration depth \citep{Asis2009,Li2016}. To further highlight the structural changes in the samples by the increased needle-punch intensity, we calculate the fibre orientation tensor $T_{ij}(z)$ and plot the $T_{zz}$ component in Figure~\ref{Fig:PoresOrientation}c. This reveals that the $z$-directional orientation of the fibres in the upper batt fibre layer is increased with increased needle-punch intensity. However, these effects seem to be limited by the used base layer structure, as sample~Low-NPI has a higher $T_{zz}$ value at a higher $z$-coordinate, which might be a fragment of the coarse base layer structure of the individual sample.

To further characterise the samples and the influence of the needle-punch intensity, we calculate the sample overall porosity $\phi_0$, the mean geodesic tortuosity vector $\vec{\bar{\tau}}$ and the single-phase permeability tensor $K_{ij}$ based on the commercial software GeoDict \citep{Wiegmann2007, Becker2021}. The results are listed in Table~\ref{tab:sampleClassification}.

\begin{table}[htbp]
\newcolumntype{C}[1]{>{\centering\arraybackslash}m{#1}} 
\renewcommand\cellalign{lc} 
\renewcommand\cellgape{\Gape[4pt]}
\caption{Classification of the sample structure based on the X-ray µCT measurements and evaluations of the structure.}
\label{tab:sampleClassification}

\begin{tabular}{l *{3}{C{3cm}}} 
\toprule
& Ref & High-NPI & Low-NPI\\
\midrule
Porosity $\phi_0$ &
0.495 & 0.496 & 0.550\\

\makecell[l]{Permeability $K_{ij}$\\ \footnotesize $[10^{-11}\,\mathrm{m}^2]$} &
$K_{ij}=\begin{bmatrix}
  6.15 & \cdots & \cdots\\
  \cdots & 6.07 & \cdots\\
  \cdots & \cdots & 2.51
\end{bmatrix}$ &
$K_{ij}=\begin{bmatrix}
  2.31 & \cdots & \cdots\\
  \cdots & 2.31 & \cdots\\
  \cdots & \cdots & 1.44
\end{bmatrix}$ &
$K_{ij}=\begin{bmatrix}
  3.68 & \cdots & \cdots\\
  \cdots & 3.65 & \cdots\\
  \cdots & \cdots & 2.20
\end{bmatrix}$\\
\addlinespace[1.5ex]

Mean tortuosity $\vec{\bar{\tau}}$ &
$\vec{\bar{\tau}}=\begin{pmatrix} 1.040 \\ 1.042 \\ 1.075 \end{pmatrix}$ &
$\vec{\bar{\tau}}=\begin{pmatrix} 1.050 \\ 1.054 \\ 1.068 \end{pmatrix}$ &
$\vec{\bar{\tau}}=\begin{pmatrix} 1.043 \\ 1.047 \\ 1.056 \end{pmatrix}$ \\
\bottomrule
\end{tabular}
\end{table}

Comparing samples~High-NPI and~Low-NPI, increasing the needle-punch intensity densifies the network as indicated by a reduction in porosity from $\phi_{0,Low} = 0.550$ to $\phi_{0,High} = 0.497$, consistent with the observed decrease in fabric height and bulk porosity $\phi_B$ (see Table~\ref{tab:sampleOverview}). The densification with increasing needle-punch intensity also increases the network complexity, reflected by higher mean tortuosity in sample~High-NPI. Consequently, the principal permeability components are lower in sample~High-NPI than in sample~Low-NPI. However, the ratio of the $z$-directional permeability to the in-plane permeability, $K_{zz}/K_{ip}$ (with $K_{ip} = \sqrt{K_{xx}^2 + K_{yy}^2}$), is higher for sample~High-NPI. This suggests enhanced liquid transport in the $z$-direction, likely due to fibre reorientation induced by needle-punching, which aligns fibres more strongly along the $z$-axis.

\section{Methods}
\subsection{Synchrotron Setup}
Time-resolved analysis of the $z$-directional liquid distribution in fibre networks is inherently challenging due to the opacity of the material. Conventional approaches using tabletop µCT~systems \citep{Wegele2024} require rather long scan times, making them unsuitable for capturing sub-second dynamics. 

The combination of micrometer-scale sampling and sub-second temporal resolution through an optically opaque nonwoven requires short exposure times and high photon flux, which is provided by synchrotron X-ray radiation. Therefore, the time-resolved X-ray radiography experiments were performed at the ForMAX~beamline \citep{Nygard2024} at the MAX~IV Laboratory, the first diffraction-limited storage ring worldwide. 


\begin{figure}[htbp]
\centerline{\includegraphics[width=\textwidth]{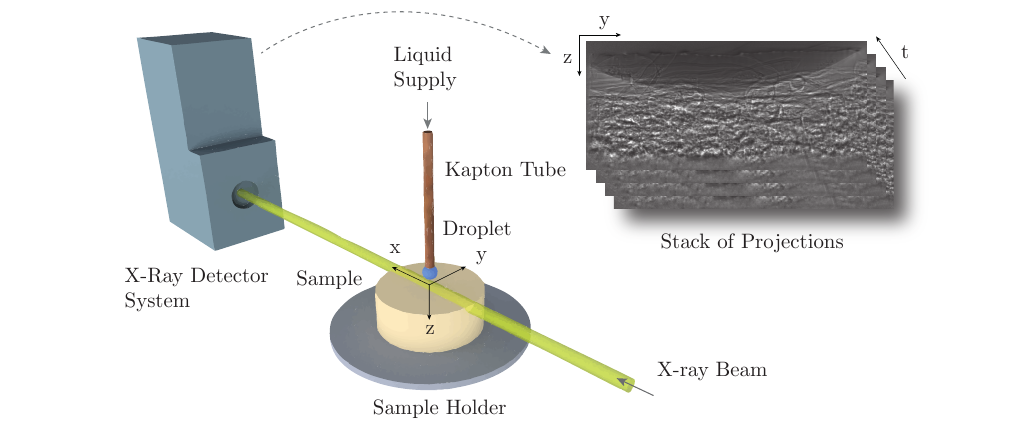}}
\caption{Schematic representation of the setup as used during the experiment. An X-ray beam is penetrating the top surface of the sample while subsequent droplets are released from the Kapton tube. The process is recorded on the X-ray detector, generating a stack of 2D-greyscale images. See Appendix~\ref{Appendix_A4} for an uncertainty analysis of the experimental setup.}
\label{Fig:Setup}
\end{figure}

In the experiment, we apply a series of droplets on the surface of an initially dry and unloaded nonwoven fibre network, progressively increasing its saturation, $\Theta$. This is performed under synchrotron X-ray illumination focused on the uppermost batt fibre layer, enabling time-resolved projection imaging of the liquid distribution within the Field of View~(FoV) using an X-ray microscope. 

The used setup is schematically shown in Figure~\ref{Fig:Setup}. The flow was driven using a syringe pump (New Era Pump System NE-4000) with a 1~ml syringe at flow rates of 1-1.5 ml/h. The liquid was dispensed through a Kapton tube (Allectra 312-Kap-Tube-07-300) with an inner diameter of 0.7~mm. The tube was mounted in a 3D-printed holder, allowing precise positioning just above the top fibre layer, ensuring droplet deposition without gravitational influence \citep{Abedsoltan2023,Davis2000, KUMAR2006}. The synchrotron beam of size 1.3~mm~×~1.5~mm illuminating the samples had an photon energy of 16.6~keV, delivering a high photon flux of $10^{15}~$ph/s/mm² at a full-harmonic bandwidth of $10^{-2}$ and was created by a multilayer monochromator~(MLM).

The collection of the X-ray beam, post-interaction with the sample, was collected with an indirect X-ray microscope. The microscope was equipped with a GaGG+ scintillator, a high-NA~5X objective and was used with an Andor Zyla~5.5 sCMOS camera. The nominal pixel size of the camera is 6.5~µm~x~6.5~µm, meaning that the effective pixel size during our measurements was 1.3~µm~x~1.3~µm. The camera was configured to operate at an acquisition framerate of up to 50~Hz. The high-temporal-resolution capabilities of ForMAX allow the deployment of the microscope with faster cameras, such as the Photron Nova~S16. Such configuration would enable us to explore faster acquisitions \citep{Olbinado2017, Asimakopoulo2024}, albeit at the expense of spatial resolution. In this work we opted for higher spatial resolution and optimized our experiments for the 50~Hz acquisition frame rate. 

To enhance X-ray contrast, we stained deionized~(DI) water with a mass concentration of $\xi$=30\% potassium iodide~(KI), as previous studies have demonstrated that this approach provides good phase contrast between air, PA6 fibres and the liquid phase \citep{Wegele2024}. Since KI slightly influences the surface tension and thereby may affect transport dynamics \citep{Clarke2002}, the surface tension of the KI-water solution is $\gamma_L = 74.0$~mN/m \citep{Hard1977}. Because the KI~solution occupies only the pore space and therefore only part of the effective beam path in projection, sufficient X-ray transmission is maintained even at high saturation states. An experimental uncertainty analysis for this experiment is provided in Appendix~\ref{Appendix_A4}. Validation of KI-based phase contrast in comparable PA6 fibre networks using forward simulations and µCT verification has been reported previously \citep{Wegele2024}.

\begin{figure}[htbp]
\centerline{\includegraphics[width=\textwidth]{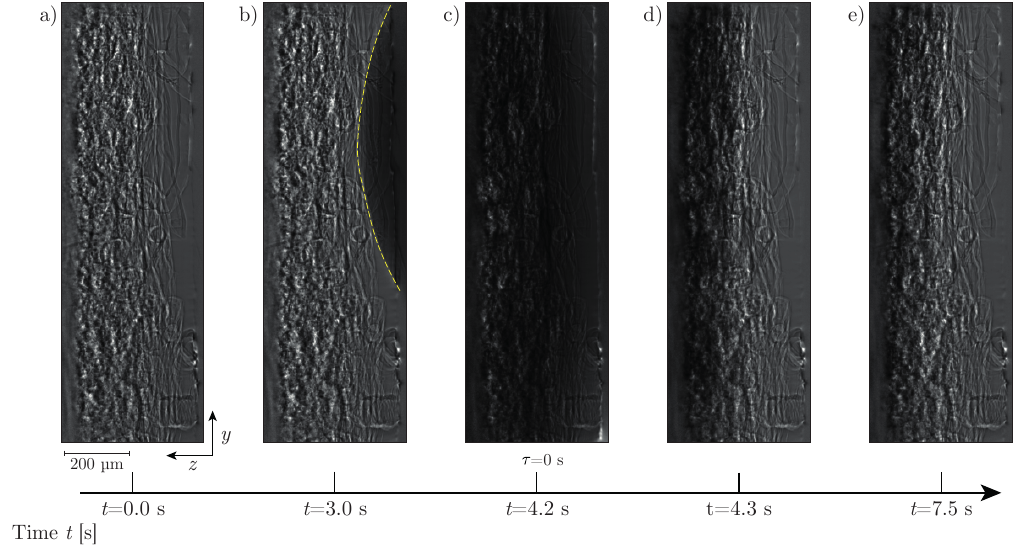}}
\caption{Flat-field corrected \citep{VanNieuwenhove2015} results as obtained on the X-ray detector for the first droplet penetrating dry sample~Ref. The droplet is coming from the right and penetrating the network visible on the left. a) Dry state, reference for dry network. b): The droplet (indicated with a dashed yellow line) is arriving. c): Droplet burst, moment of absolute saturation in top level and reference for $\tau$=0~s for later model fitting. d): Liquid transport within the network decreases the absolute saturation in the FoV. e): Static situation where liquid has redistributed just before the next droplet is released.}
\label{Fig:OriginalResults_1}
\end{figure}

The X-ray radiography provides raw greyscale images as depicted in Figure~\ref{Fig:OriginalResults_1} for the first droplet applied on sample~Ref. The approaching droplet can clearly be distinguished on the detector due to the high X-ray attenuation coefficient of the water-KI solution. Upon contact with the network, capillary forces overcome the surface tension, leading to a rapid droplet collapse and immediate in-plane and $z$-directional spreading. This results in an initial reduction in greyscale intensity within the FoV, followed by a gradual recovery as the liquid redistributes throughout the network. Once equilibrium is reached, the next droplet is applied, enabling investigation of the spreading dynamics as a function of both time and network saturation.

\begin{figure}[htbp]
\centerline{\includegraphics[width=.5\textwidth]{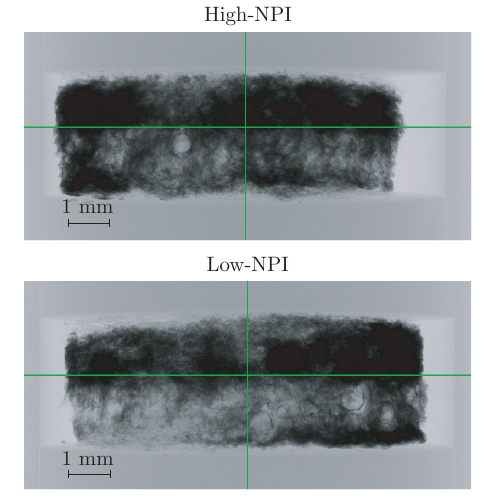}}
\caption{Static X-ray projection scan of samples~High-NPI~and~Low-NPI in the µCT after the experiment was conducted. Dark areas are the result of dried KI, indicating that the KI-stained liquid was present within these areas.}
\label{Fig4}
\end{figure}

To verify whether the liquid had penetrated the full thickness of the samples, static X-ray projection scans, as shown in Figure~\ref{Fig4}, were performed on samples~High-NPI and~Low-NPI after drying. As KI was used as a tracer, dark areas are apparent where the liquid was located and dried after the experiment. No contrast enhancement inside PA6~fibres was observed in the X-ray projection scans of the dried samples, indicating that KI residues remain associated with the liquid-filled pore space rather than the fibre material. Both samples exhibit dark regions distributed throughout the entire sample volume, indicating that the liquid reached all regions of the sample, confirming in-plane and $z$-directional penetration. Notably, the majority of the liquid accumulated in the fine pores of the uppermost fibre layer, consistent with previous observations of the static liquid distribution in similar systems \citep{Wegele2024}. These results confirm that the applied liquid is transported across the full thickness of the samples, thereby justifying analysis of $z$-directional liquid transport dynamics.

\subsection{Evaluation of network saturation}
\label{Sec:NetworkSaturation}
In our experiment, a sequence of liquid droplets is applied to an initially dry fibre network. As the network becomes wetted, individual pores are progressively filled with liquid possessing a high X-ray attenuation coefficient, displacing the air phase, which exhibits much lower attenuation. According to Beer–Lambert's law, this replacement leads to a decrease in transmitted X-ray intensity~$I$, which is captured by the detector as a corresponding greyscale value~$I_i(y,z)$ in frame~$i$ of the 2D~projection at position~$(y,z)$. Even though inverting Beer-Lambert's law would justify using a logarithmic correlation of the saturation on the measured X-ray intensity, it is state-of-the-art in X-ray projection imaging of porous media to use a linear approach that is normalized based on the X-ray intensity of the dry network \citep{Akin2003, Zhang2014,Zhang2019, Sheikhi2023}. Following that, we calculate the relative dryness $1-\Theta(y,z)$ of the network based on the measured X-ray intensities of the dry~$I_{d}$ and fully saturated network~$I_{s}$:

\begin{equation}
\label{Eq:IntensitySaturation}
   1-\Theta(y,z)=\frac{I_{s}-I(y,z)}{I_{s}-I_{d}}
\end{equation}

\begin{figure}[htbp]
\centerline{\includegraphics[width=\textwidth]{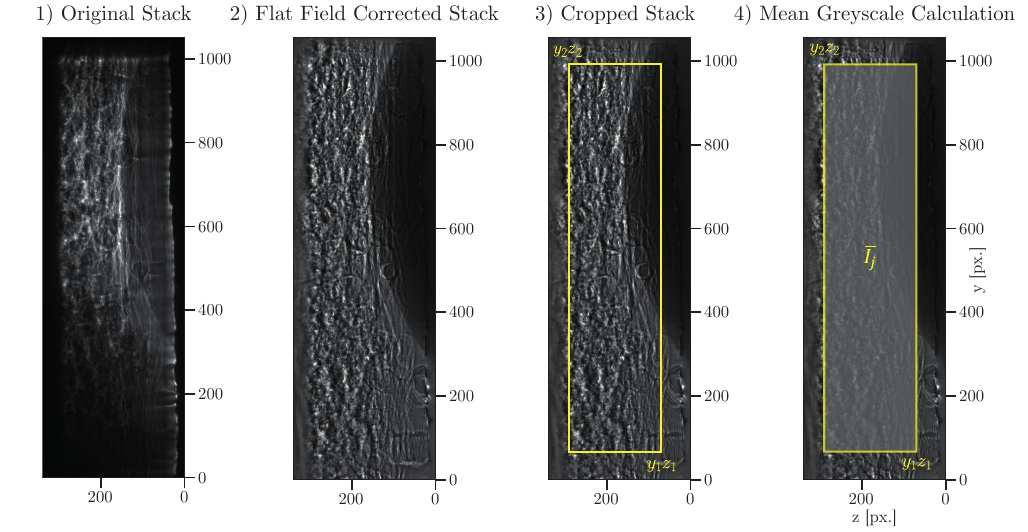}}
\caption{Evaluation algorithm to determine the mean greyscale value $\bar{I_j}$ for each frame of the processed stacks.}
\label{Fig5}
\end{figure}

Eq.~\ref{Eq:IntensitySaturation} is interpreted as a normalized, intensity-based proxy between the measured dry reference~$I_d$ and a sample-specific saturated reference~$I_s$. In Beer-Lambert form one may write $I(t) = I_s\cdot e^{[-\Delta\mu(l(t)-l_s)]}$, where $l(t)$ is the effective KI-solution thickness along the beam. For moderate variations around the reference state, one can assume $e^{[-\Delta\mu(l(t)-l_s)]}\approx 1-\Delta\mu(l(t)-l_s)$ (first-order approximation), motivating the linear mapping in Eq.~\ref{Eq:IntensitySaturation}. A detailed derivation is provided in Appendix~\ref{Appendix:Linearization}.

To assess network saturation in the uppermost layer as a function of time, we follow the workflow as shown in Figure~\ref{Fig5}. In an initial preprocessing step, we perform a conventional flat-field correction \citep{VanNieuwenhove2015} to the image stack using 100~flat-field images acquired prior to the experiment. To account for variations in sample height and placement and to reduce the sensitivity to local artifacts, we define specific overlaying cropped regions $j = (y_1, y_2, z_1, z_2)$ for further evaluation. The average intensity $\bar{I}_j$ is then computed within each cropped region $j$ across the image stack. To normalise the average intensity on the dry state as indicated in Eq.\ref{Eq:IntensitySaturation}, the average intensity of the dry network $\bar{I}_d$ (measured before the first droplet is applied) is used as a reference. The frame exhibiting the lowest average intensity in each droplet release is identified as the moment of droplet burst, which marks the onset of liquid redistribution and defines the relative droplet time $\tau$.

\begin{figure}[htbp]
\centerline{\includegraphics[width=.5\textwidth]{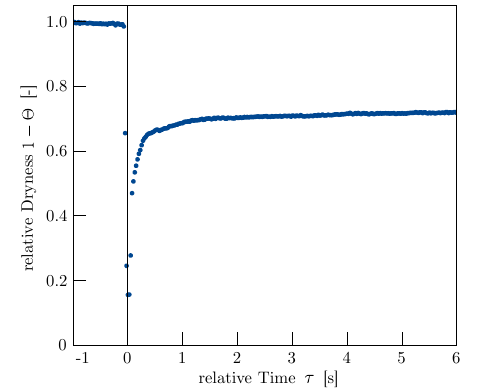}}
\caption{Principal result of evaluating the network dryness for the first droplet on sample~Ref at crop $j=(25,190,20,1000)$ for a framerate of 40~Hz. During droplet arrival, the value for $1-\Theta$ is gradually reduced, ending with the droplet burst at $\tau=0$. From that point on, a redistribution of liquid within the networks takes place, reducing the local saturation within the FoV.}
\label{Fig:PrincipalSaturation}
\end{figure}

Figure~\ref{Fig:PrincipalSaturation} shows the evolution of $\bar{I}_j$ for the first droplet applied to sample~Ref. Using Eq.~\ref{Eq:IntensitySaturation}, we can calculate the relative dryness $1-\Theta$ out of the average intensity $\bar{I}_j$. Starting from an initially dry network, the relative dryness is reduced by the arriving droplet. At the droplet burst, a minimum value is reached before capillary forces cause a redistribution of the liquid in the whole network out of the FoV, resulting in an increasing value in relative dryness. The process of liquid redistribution has a pronounced nonlinearity and levels off with increasing time $\tau$.

Because droplet deposition irreversibly changes the saturation state and wetting history, each sample was measured only once. To account for potential sensitivity to sample placement and local heterogeneity, intensity evaluation was repeated for four distinct crop regions and the mean value was used thereafter (see crop coordinates in Appendix~\ref{Sec:App1}). The sample was not translated during acquisition because the time-resolved analysis requires tracking a fixed region during burst and redistribution. Since $\mathrm{FoV}\gg d_{\mathrm{batt}}$ and the signal is integrated through the sample thickness in projection imaging, modest in-plane shifts are not expected to qualitatively affect the saturation-dependent trends.

\subsection{Derivation of Liquid Transport Function}
\label{Sec:SpreadFunction}

\begin{figure}[htbp]
\centerline{\includegraphics[width=.5\textwidth]{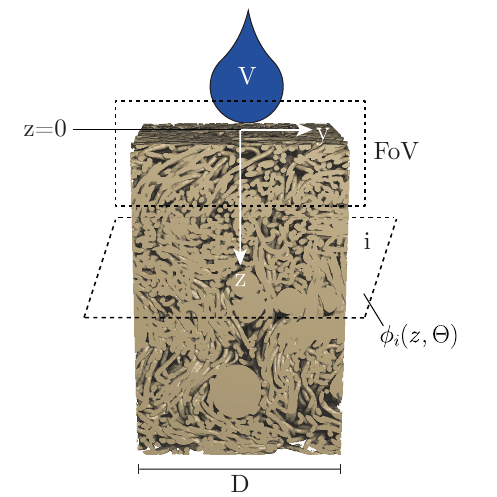}}
\caption{Model assumption to derive a physical description of the experiment. A droplet of volume~V is applied on a circular fibre network with diameter~D, from which we see the 2D~projection within the FoV only at the top layer. The remaining porosity can be a function of the $z$-direction and the saturation.}
\label{Fig:ZSpreadSchmema}
\end{figure}

In the experimental procedure, successive droplets are applied to the initially dry fibre network and the resulting liquid redistribution is assumed from changes in greyscale intensity, driven by the contrast in X-ray attenuation between the liquid and the surrounding air. As illustrated in Figure~\ref{Fig:ZSpreadSchmema}, the FoV is considerably smaller than the sample. Therefore, a model is required to assess the liquid transport dynamics within the whole sample. Despite the known limitations, we assume that the liquid penetration depth $z(\tau)$ into the network can be reasonably approximated by the classical Lucas–Washburn equation, as capillary forces govern the droplet distribution:

\begin{equation}
\label{Eqn:LW-Equation}
    z(\tau)=\sqrt{\frac{\gamma_L \cdot cos(\theta) \cdot r}{2\mu}\cdot \tau} 
\end{equation}

Within Eq.~\ref{Eqn:LW-Equation}, $\theta$ denotes for the solid-liquid contact angle while $\mu$ is the dynamic viscosity of the liquid. After the droplet of volume $V=Q\cdot \Delta t$ was released, it is assumed that it immediately spreads within the plane of the network and travels down in the $z$-direction a distance $z(\tau)$. The invading volume $V_{inv}$ of the droplet in a network of remaining porosity $\phi(z,\Theta)$ can then be described with the diameter $D$ of the sample:

\begin{equation}
    V_{inv}(\tau)=\int_{0}^{z(\tau)} \phi(z,\Theta)\cdot\frac{\pi}{4}\cdot D^2 \,dz
\end{equation}

The liquid volume $V_{z=0}(\tau)$ at the top surface of the sample is continuously reduced as the liquid invades the network. It can be calculated by:
\begin{equation}
    V_{z=0}(\tau)=V-V_{inv}(\tau)=V-\int_{0}^{z(\tau)} \phi(z,\Theta)\cdot\frac{\pi}{4}\cdot D^2 \,dz
\end{equation}

Assuming that the porosity is evenly distributed within the network ($\phi(z)$=const.), the integral is simplified to:
\begin{equation}
    V_{z=0}(\tau)=V-\bar\phi(\Theta)\cdot \frac{\pi}{4}\cdot D^2\cdot z(\tau)
\end{equation}

The liquid content $q_{z=0}(\tau)$ in the uppermost fibre layer can then be calculated using Eq.~\ref{Eqn:LW-Equation}:
\begin{equation}
\label{Eqn:PhysicalSpreadEquation}
    q_{z=0}(\tau)=1-\frac{1}{V}\cdot \bar\phi(\Theta)\cdot \frac{\pi}{4}\cdot D^2\cdot \sqrt{\frac{\gamma_L \cdot cos(\theta) \cdot r}{2\mu}\cdot \tau}
\end{equation}

As the recorded greyscale is proportional to the saturation level in the uppermost batt fibre layer, we are using the following approach function to fit the recorded experimental data of the liquid content in the network, introducing the penetration constant~$C$ that can be obtained for a single droplet spread in the network:

\begin{equation}
\label{Eqn:FitFunction}
    q_{z=0}(\tau)=\frac{1}{1-C\cdot\sqrt{\tau}}
\end{equation}

Please note that the inverse formulation is a consequence of evaluating the greyscale of the images, which is inversely proportional to the liquid content (see Eq.~\ref{Eq:IntensitySaturation}). The function fit as such was applied on the first 200 obtained frames per sample, which corresponds to $\Delta\tau=5$~s (40~Hz, sample Ref) or $\Delta\tau=4$~s (50~Hz, samples High-NPI and Low-NPI) of redistribution dynamics.

In the Lucas–Washburn model, the flow front progresses as $z(\tau)\propto \sqrt{\tau}$, whereas in the more advanced Bosanquet model, the penetration depth scales linearly with time, $z(\tau)\propto \tau$. As shown in Figure~\ref{Fig:PrincipalSaturation}, our results exhibit a distinctly nonlinear saturation behaviour within the FoV, supporting the use of a Lucas–Washburn-based formulation to fit the experimental data.

\subsection{Introduction of a $z$-directional liquid transport index~$T_z(\Theta)$}
Fitting the experimental data to Eq.~\ref{Eqn:FitFunction} yields the saturation-dependent parameter~$C(\Theta)$. However, the underlying physical model in Eq.~\ref{Eqn:PhysicalSpreadEquation} includes the porosity~$\phi(z,\Theta)$, which decreases during subsequent droplet release. By knowing the initial porosity~$\phi(z,0)$ based on dry µCT scans as well as the volumes $V_i$ of all individual droplets being applied on the network, we can determine the average porosity~$\bar{\phi}(\Theta)$ for each saturation state~$\Theta$, with the simplifying assumption that the pores are equally distributed in the fibre networks. Furthermore, several parameters like the solid-liquid surface tension~$\gamma_L$, dynamic viscosity~$\mu$ and the sample diameter~$D$ in Eq.~\ref{Eqn:PhysicalSpreadEquation} are known a priori and remain constant during the experiment. Hence, we decompose the penetration constant~$C(\Theta)$ into two components: a term~$A(\Theta)$, which encapsulates all known parameters and a term~$T(\Theta)$, referred to as the transport index, which contains the remaining unknowns:
\begin{equation}
    A(\Theta)=\frac{1}{V}\cdot \bar\phi(\Theta)\cdot\frac{\pi}{4}\cdot D^2\cdot\sqrt{\frac{\gamma_L}{2\mu}}
\end{equation}
\begin{equation}
    T(\Theta)=\sqrt{r(\Theta)\cdot cos(\theta)}
\end{equation}
While the classical Lucas–Washburn equation assumes a constant capillary radius~$r$, this assumption does not hold for partly-saturated fibre networks where effective pore sizes increase with saturation as larger pores become accessible to the liquid phase \citep{vanGenuchten1980, Tavangarrad2019}. As we do not know the capillary pressure function~$p_c(\Theta)$ of the individual samples, we assume it to be unknown. Furthermore, the contact angle~$\theta$ is influenced by the velocity of the liquid front due to dissipation at the contact line \citep{Voinov1977, Cox1986}, introducing further uncertainty into the formulation. Consequently, Eq.~\ref{Eqn:PhysicalSpreadEquation} is reformulated:

\begin{equation}
    q_{z=0}(\tau)=1-A(\Theta)\cdot T(\Theta) \cdot \sqrt{\tau} 
\end{equation}

This allows us to calculate the transport index for different saturation levels~$T(\Theta)$ using the parameter~$A(\Theta)$ and the fitted parameter~$C(\Theta)$:
\begin{equation}
    T(\Theta)=C(\Theta)/A(\Theta)
\end{equation}

In contrast to Figure~\ref{Fig:ZSpreadSchmema}, the liquid spreads in both the in-plane and $z$-directions. While directional permeability is formally defined within the framework of pressure-driven flow and describes the pressure–velocity relationship, it also inherently reflects the geometric anisotropy and structural characteristics of the porous network. Therefore, it provides meaningful information about the preferential pathways available for capillary-driven flow. Based on this, we define an anisotropy factor~$\alpha$ to capture the directional flow resistance, as the ratio of the $z$-directional to in-plane single-phase permeabilities:

\begin{equation}
   \alpha=\frac{K_z}{K_{ip}} = \frac{K_{zz}}{\sqrt{K_{xx}^2 + K_{yy}^2}}
   \label{eq:Anisotropy}
\end{equation}

With that, we can calculate the $z$-directional transport index $T_z(\Theta)$, which is a measure of the liquid transport dynamics as a function of the saturation state:
\begin{equation}
\label{Eqn:zLiquidSpread}
    T_z(\Theta)=T(\Theta)\cdot \alpha
\end{equation}

As high values of~$T_z(\Theta)$ will render a high flow length~$z$ at identical times~$\tau$ in Eq.~\ref{Eqn:PhysicalSpreadEquation}, we can interpret~$T_z(\Theta)$ as a number, denoting how fast the liquid distributes in the $z$-direction within the samples. Hence, it is seen as a performance indicator for the fibre network, which is derived from fitting the Lucas-Washburn-based model to the experimental data.

While the transport index~$T_z(\Theta)$ is not equivalent to the relative permeability in a strict sense, it captures similar physical phenomena. Specifically, it reflects the ease and speed with which liquid is transported through the network under capillary forces. Unlike permeability, which describes steady-state flow under an applied pressure gradient, it quantifies transient capillary-driven transport dynamics influenced by pore structure, orientation and evolving saturation. As such, it serves as a useful proxy for assessing directional flow performance in unsaturated fibrous media in addition to relative permeability.

\section{Results and Discussion}
Using the derived model to describe liquid transport dynamics, we conducted droplet experiments on sample~Ref to investigate the fundamental $z$-directional transport behaviour over a wide range of saturation states. Data for sample~Ref were acquired at a frame rate of 40~Hz. To assess the influence of the needle-punch intensity, the experiment was repeated on samples~High-NPI~and~Low-NPI at a frame rate of 50~Hz.

\subsection{Principal behaviour}
To investigate the principal $z$-directional liquid transport behaviour as function of time and saturation, we applied 15~subsequent droplets at a flow rate of 1~ml/h to sample~Ref and investigated the mean X-ray intensity, which corresponds to the mean saturation within the uppermost layer of the network. The determination of $C$ with the function fit for crop~1=(25,~190,~20,~1000) according to Eq.~\ref{Eqn:FitFunction} is visible in Appendix~\ref{Sec:App2}, while Appendix~\ref{Sec:App1} provides an overview on all crop coordinates, individual C-values and associated standard deviations. 

\begin{figure}[htbp]
\centerline{\includegraphics[width=\textwidth]{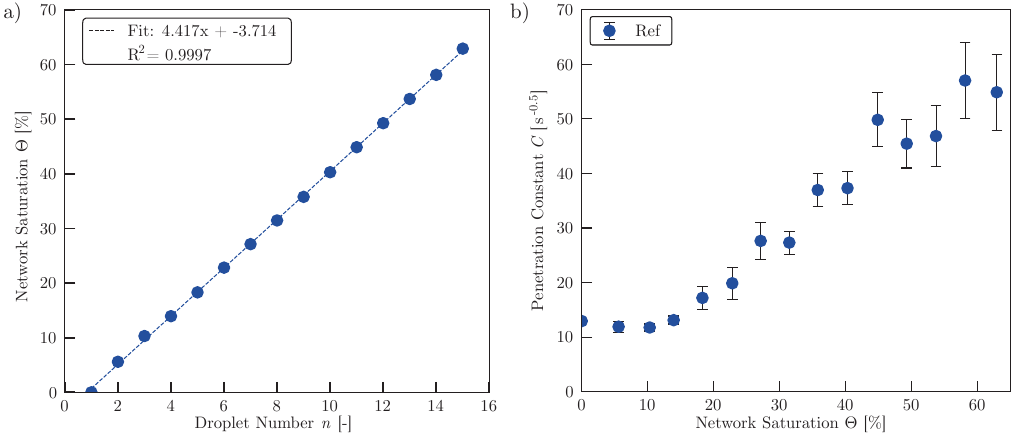}}
\caption{a) Saturation of sample~Ref after each droplet penetrated the sample. Droplet volume was calculated by using the applied flow rate and the time between the droplet collapse points. b) Mean value of the penetration constant~$C$ as a function of the network saturation. $C$~was determined by evaluating the results of four different crops and calculating the mean value. Error bars depict the standard deviation of the individual values of~$C$. Mean values and standard deviations are reported in Appendix~\ref{Sec:App1})}
\label{Fig:Sat_C_SampleA}
\end{figure}

Across all droplets, the mean greyscale signal exhibits a consistently nonlinear temporal response and the initial saturation before each droplet release increases progressively. As the number of applied droplets increases, the liquid redistributes more rapidly, with the mean X-ray intensity reaching an asymptotic value more quickly. The fitted model accurately captures this dynamic behaviour, allowing the determination of the transport dynamics parameter~$C$ for each liquid redistribution process. Given the known initial porosity~$\phi_0$, the saturation state~$\Theta$ after each droplet can be calculated as shown in Figure~\ref{Fig:Sat_C_SampleA}a. Each droplet increases the saturation by approximately 4.4\%, reaching a final saturation of 67.1\% after the last droplet. The near-linear relationship between droplet number and saturation confirms the repeatability and consistency of the experimental procedure. Figure~\ref{Fig:Sat_C_SampleA}b shows the evolution of the penetration constant~$C(\Theta)$ with increasing saturation. After an initial decrease from $C(0)$=12.95 to $C(10.3)$=11.76, $C$ rises sharply, reaching a value of $C(58.1)$=57.04. The standard deviation of the mean values determined from the individual crops also increases with higher saturation. These results suggest that liquid redistribution accelerates with increasing saturation, as the equilibrium moisture content in the uppermost layer is reached faster at higher values of~$C$.

\begin{figure}[htbp]
\centerline{\includegraphics[width=.5\textwidth]{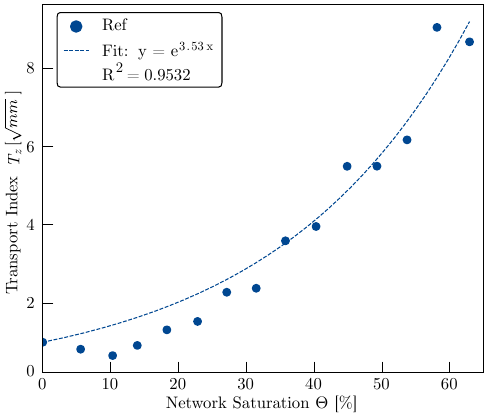}}
\caption{$z$-directional transport index $T_z$ normalized to $T_z (\Theta=0)$. An exponential function is fitted to the values, allowing to assess the exponent~$a$.}
\label{Fig:Res_TIz_a}
\end{figure}

To assess the $z$-directional liquid transport as a function of the network saturation, we use the mean values for $C(\Theta)$ and calculate the $z$-directional transport index $T_z(\Theta)$ using the anisotropy factor $\alpha_{Ref}=0.291$, which is calculated using the permeability values $K_{ii}^s$ from Table~\ref{tab:sampleClassification} in Eq.~\ref{eq:Anisotropy}. The results are shown in Figure~\ref{Fig:Res_TIz_a}. The data reveal a steady increase of $T_z(\Theta)$ with saturation, following a brief initial decline. Since $T_z(\Theta)$ reflects the rate of liquid redistribution, this trend indicates that liquid transport through the network is faster as the saturation increases. The acceleration is particularly pronounced at higher saturation levels.

Given that relative permeability in fibrous media is often modeled as an exponential function of saturation, we fit an exponential function of the kind $T_z(\Theta)=e^{a\cdot \Theta}$ to the experimental data for $T_z(\Theta)$, yielding an exponent of $a_{Ref}=3.53$. This is consistent with previous findings on the relative in-plane permeability of nonwovens with straight fibres \citep{ASHARI2010} ($3.3<a<4.1$) and needle-punched fibre networks \citep{Landeryou2005} ($a=3$), thereby extending those findings to the $z$-direction. 

Notably, for saturation levels below $\Theta$=32\%, the observed $T_z$ values fall below the fitted exponential trend. This deviation is consistent with the presence of a percolation threshold, as previously reported for fibre networks \citep{Landeryou2005} and soils \citep{Haverkamp1997}, where limited connectivity in the pore space suppresses liquid transport below a critical saturation level. Furthermore, the dynamics of liquid spreading are strongly influenced by the wetting state of the microstructure \citep{Kim2013}. On pre-wetted surfaces, where a liquid film is already present, the spreading occurs significantly faster compared to initially dry conditions \citep{Fraaije1989}. This indicates that an initial liquid volume is required to wet the fibre surfaces, after which the liquid advances more rapidly along the wetted pathways.
As the exponential function allows for modelling the liquid transport index $T_z(\Theta)$ with high accuracy, we conclude that this modelling function will also be valid for saturation states beyond the range investigated in the underlying study.

\subsection{Influence of the needle-punch intensity}
\begin{figure}[htbp]
\centerline{\includegraphics[width=\textwidth]{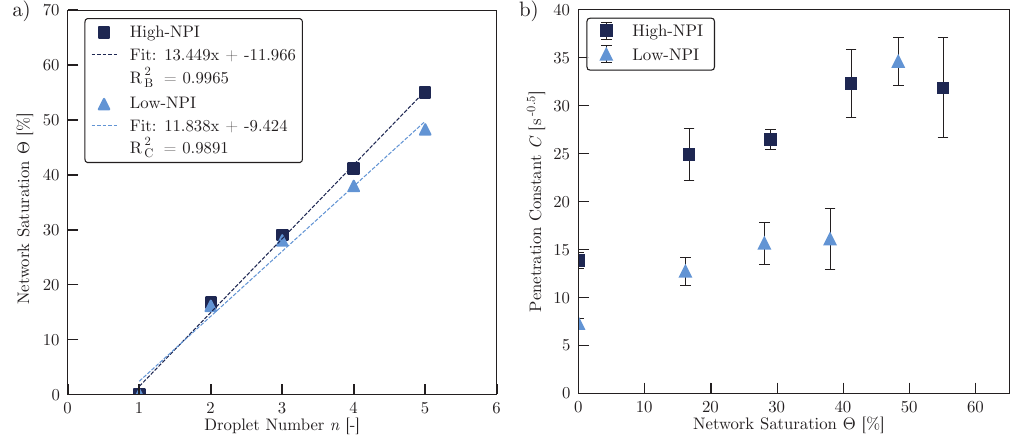}}
\caption{a) Saturation of samples~High-NPI and~Low-NPI after each droplet penetrated the sample. Droplet volume was calculated by using the applied flow rate and the time between the droplet collapse points. b) Mean value of the penetration constant~$C$ as a function of the network saturation. $C$~was determined by evaluating the results of four different crops and calculating the mean value. Error bars depict the standard deviation of the individual values of~$C$. Mean values and standard deviations are reported in Appendix~\ref{Sec:App1})}
\label{Fig:sat_2}
\end{figure}

Having established the principal behaviour of $z$-directional liquid transport in needle-punched fibre networks, we now investigate the effect of the needle-punch intensity by repeating the experiment with samples~High-NPI and~Low-NPI, which differ in their needle-punch intensity. As we again expect an exponential behaviour of $T_z(\Theta)$, we apply five droplets at an increased flow rate of $Q=$1.5~µl/h to reach similar saturation conditions within shorter times. Based on the known initial porosity of each sample, the corresponding saturation values were calculated after each droplet application, as shown in Figure~\ref{Fig:sat_2}a.

Both samples exhibit similar saturation trends, with final saturation levels of $\Theta_{High}$=66.6\% for sample~High-NPI and $\Theta_{Low}$=64.2\% for sample~Low-NPI after the fifth droplet. The slightly higher saturation in sample~High-NPI is attributed to its lower initial porosity $\phi_{0,{High}}$, which results in a higher saturation gain per unit of added liquid. The values of~$C$ for samples~High-NPI and~Low-NPI were determined using the fitting algorithm described in Section~\ref{Sec:SpreadFunction}, applied to four independent crop regions per sample (see Appendix~\ref{Sec:App1} for crop coordinates, individual C-values and associated standard deviations). The mean values of~$C$ as a function of network saturation~$\Theta$ are shown in Figure~\ref{Fig:sat_2}b. 

For both samples, $C$ increases consistently with saturation, except for the final droplet, where the trend deviates and $C_{Low}$ exceeds $C_{High}$. Overall, $C_{High}$ values are higher than $C_{Low}$ for droplets 1-4, whereas the final droplet is the only point where this ordering reverses. It is important to note that each of the samples High-NPI and Low-NPI was measured only once, without physical replicates. While the use of multiple crop regions helps to mitigate local variability and measurement noise (indicated by the error bars in Fig.~\ref{Fig:sat_2}b), these pseudo-replicates do not allow a statistical significance assessment of differences between High-NPI and Low-NPI at the sample level. Notably, for droplets 1-4 the difference in the mean value of $C$ between High-NPI and Low-NPI exceeds the crop-to-crop variability, whereas the final droplet is the only point where this is not the case.

Unlike sample~Ref, samples~High-NPI and~Low-NPI do not exhibit an initial decline in~$C$ at low saturation. This difference may result from the fact that higher saturation levels are reached immediately after the first droplet, due to the higher volume of each droplet used during testing samples~High-NPI~and~Low-NPI.

\begin{figure}[htbp]
\centerline{\includegraphics[width=.5\textwidth]{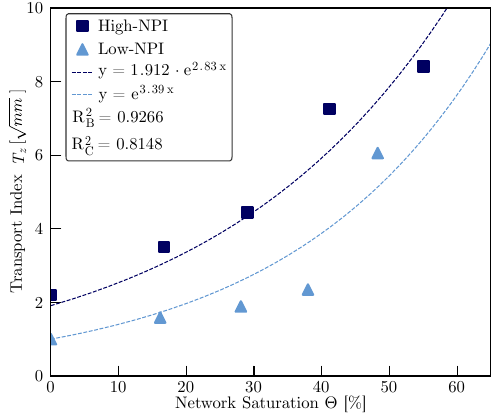}}
\caption{Normalized values for $T_z$ for samples~High-NPI and Low-NPI with different needling density as a function of the network saturation. The values are normalize to~$T_{z,Low}(\Theta=0)$, as this was the lowest transport index observed for samples~High-NPI and Low-NPI.}
\label{Fig:k_rel}
\end{figure}

The $z$-directional transport index~$T_z(\Theta)$ for samples with varying needle-punch intensity is presented in Figure~\ref{Fig:k_rel}. For comparative reasons, the~$T_z(\Theta)$ for samples~High-NPI and~Low-NPI was normalized to the dry-state value of sample~Low-NPI ($T_z(\Theta=0)$), which exhibited the lowest initial transport index. As observed previously, $T_z$~increases with saturation and is well described by an exponential function. In the dry state, sample~High-NPI shows a transport index approximately 91\% higher than that of sample~Low-NPI. Fitting exponential curves to the data yields exponents $a_{High}=2.52$ and  $a_{Low}=3.39$ for samples~High-NPI and~Low-NPI, respectively. These results indicate that the $z$-directional liquid transport in sample~Low-NPI is more sensitive to increasing network saturation. However, the initial offset of $T_{z,High}(\Theta=0)$ causes that sample~High-NPI will provide a higher $z$-directional transport index until full network saturation, which can be predicted by extrapolation of the fitted functions. This finding suggests that increased needle-punch intensity enhances $z$-directional liquid transport, despite reducing the overall permeability in all principal directions. We attribute this behaviour to fibre reorientation induced by the needle-punch process: fibres originally aligned in-plane are redistributed towards the $z$-direction. Since liquid flow along the fibre axis encounters less resistance than flow perpendicular to it, this structural change creates preferred pathways for $z$-directional transport. This interpretation is supported by the observed differences in the permeability anisotropy factor~$\alpha$, which is higher for sample~High-NPI~($\alpha_{High}=$0.4405) than for sample~Low-NPI~($\alpha_{Low}=$0.4251). These results underscore the critical role of needle-punching not only in ensuring structural cohesion but also in defining the functional transport properties of nonwoven fibre networks.

\section{Conclusion}
\label{sec:Sec_Conclusion}
This study demonstrates that time-resolved synchrotron X-ray radiography is a powerful tool for investigating the dynamics of liquid transport in nonwoven fibre networks, particularly in the $z$-direction. The ability to visualize in-situ liquid distribution through the sample thickness with high spatial and temporal resolution offers new insights into the mechanisms governing capillary-driven flow in fibrous materials.
Despite the known limitations of the Lucas–Washburn equation, we developed a simplified physical model based on it that adequately captures the observed capillary-driven liquid transport behaviour. When combined with static µCT~scans of the dry network, this framework enabled us to define a $z$-directional liquid transport index,~$T_z(\Theta)$, which characterizes the dynamic liquid transport as a function of saturation~$\Theta$.
Model fitting revealed that $T_z$~increases exponentially with network saturation, consistent with prior findings on in-plane relative permeability in fibrous media. Furthermore, by comparing fibre networks that differ only in needle-punch intensity, we demonstrated the critical role of network structure in facilitating $z$-directional transport. Although increased needling intensity reduces single-phase permeability by densifying the network, it simultaneously promotes $z$-oriented fibre alignment. This reorientation creates preferential flow paths in the thickness direction, substantially enhancing $z$-directional liquid transport in partly-saturated conditions. These findings highlight that needle-punching is not only a necessary manufacturing step for achieving mechanical cohesion but also a key design parameter for tailoring the liquid transport properties of nonwoven materials. Needling parameters such as intensity and depth should therefore be optimized not only for structural performance but also to meet specific liquid transport requirements in targeted applications.

However, a significant limitation of this study is that each condition was tested using only a single sample. Given the inherent structural variability in nonwoven fibre networks, this restricts the statistical robustness of our conclusions. To increase confidence in the observed trends, future investigations should include independent replicate measurements across multiple samples from the same material batch. In addition, benchmark experiments on reference media with well-defined pore structures, together with repeated measurements, would enable a complete uncertainty analysis by separating instrumental and processing variability from sample-to-sample variability. While our results reveal consistent and interpretable trends, the scope of the experimental study remains limited, particularly in the number of needle-punching conditions tested. Future work should expand the parameter space by systematically varying needling depth, intensity and network architecture, while incorporating replicates, to develop a more comprehensive and statistically validated understanding of how structural modifications influence capillary-driven liquid transport in porous fibrous media.

In addition, future studies could explicitly investigate the droplet stability (burst) threshold as a function of the network saturation $\Theta$, for example, by relating the Laplace pressure at burst to an effective, saturation-dependent capillary entry pressure. Such an analysis would provide further insight into capillary entry effects, wetting hysteresis and the role of pre-wetting on the onset of rapid penetration into fibrous networks.

\backmatter

\bmhead{Acknowledgements}
The authors would like to thank Richard Westerholz for his support with tomographic measurements of the samples. Kjell Karlsson and Johan Malmqvist are thanked for their support in manufacturing the samples. We acknowledge the MAX~IV~Laboratory for beamtime on the ForMAX~beamline under proposals~20231192 and~20240748. Research conducted at MAX~IV, a Swedish national user facility, is supported by Vetenskapsrådet~(Swedish Research Council, VR) under contract~2018-07152, Vinnova~(Swedish Governmental Agency for Innovation Systems) under contract~2018-04969 and Formas under contract~2019-02496. This work received funding and support from the ERC-2020-STG, 3DX-FLASH~(948426).

\bmhead{Author contribution}
Conceptualization:~[PW, TR, LDS], Methodology:~[PW, PVP, EMA, ZY, JKR], Data Curation~[ZY, PW], Formal Analysis:~[PW], Resources:~[KN, ZY, JKR, EMA, PVP], Investigation:~[PW, ZY, JT, AG, JKR, EMA, PVP], Visualization:~[PW], Supervision:~[LDS, TR, PVP], Writing~-~original~draft:~[PW], Writing~-~reviewing~\&~editing:~All~authors

\bmhead{Data availability}
The data that support the findings of this study are available upon reasonable request from the authors.

\section*{Declarations}

\bmhead{Conflict of interest}
Patrick Wegele reports that financial support and equipment were provided by J.M.~Voith~SE~\&~Co.~KG. Patrick Wegele reports a relationship with J.M.~Voith~SE~\&~ Co.~KG, including employment. The other authors declare that they have no competing interests.

\bmhead{Ethical approval}
Not applicable.
\newpage

\begin{appendices}
\section{Experimental uncertainty analysis}
\label{Sec:App0}
\subsection{Porosity $\phi_{0,i}$}
The porosity of the individual samples is determined based on the acquired tomograms. After segmentation using the Otsu method \citep{Otsu1979}, the porosity $\phi_{0,i}$ can be determined by voxel counting. To assess the consistency of the resulting porosity values, they are compared to calculated values of the bulk porosity $\phi_{B,i}$ of the networks, where the samples had been cut out. The discrepancy $\Delta\phi=\phi_{B,i}-\phi_{0,i}$ in the µCT-based calculation of the porosity $\phi_{0,i}$ compared to the gravimetrically calculated bulk porosity $\phi_{B,i}$ are within 0.4-2.7\% (absolute deviation), supporting that segmentation-based porosity is representative. The apparent differences are attributed to inhomogeneities of the detailed network structure of the samples compared to the individual networks, uncertainties in the thickness and grammage measurements of the individual networks and partial-volume effects at 6~µm voxels.

\subsection{Pore Size Distribution}
Based on the µCT scans, the pore size distribution in the samples can be determined. It is calculated using a granulometry approach as described by \cite{Silin2006} and implemented by \cite{Schulz2007}, resulting in a pore size of a morphological inscribed-sphere equivalent diameter. To avoid resolution-limited bias, pore sizes below 25~µm are not interpreted. Given a voxel size of 6~µm, 25~µm corresponds to 4.2~voxels. Furthermore, the minimum fiber diameter is 35~µm (5.8~voxels), such that fiber geometry is resolved and partial-volume effects are reduced. Remaining uncertainty primarily affects the lower tail near the cutoff and does not affect the qualitative comparisons discussed here.

\subsection{Fibre Orientation $T_{zz}(z)$}
$T_{zz}(z)$ is calculated using a Star Length Distribution~(SLD) approach implemented in the commercial software GeoDict \citep{Becker2021}, which determines a local~(voxel-based) orientation tensor from chord lengths evaluated along a discrete set of directions and subsequently averages these tensors within each thickness layer. In our case, fibers are resolved by multiple voxels across their diameter (minimum fiber diameter $\gg$ voxel size), which limits voxelization and partial-volume bias in the orientation estimate. Moreover, each reported value $T_{zz,i}$ represents an average over the full cross-section of the corresponding layer, making the result insensitive to local image noise. The remaining uncertainty is therefore dominated by systematic contributions, in particular the finite directional sampling in the SLD evaluation, segmentation quality at fiber boundaries and the representativeness of the selected µCT sub-volume. These effects may slightly shift the absolute values of $T_{zz,i}$, but they are not expected to alter the qualitative through-thickness trends and sample-to-sample comparisons discussed in this work. As the evaluation is performed per thickness layer (without lateral subdivision), the reported $T_{zz}(z)$ represents the mean orientation of the scanned region of interest and does not quantify lateral variability.

\subsection{Time-resolved radiography experiment}
\label{Appendix_A4}
Within the time-resolved radiography experiment, the main sources of uncertainty arise from variations in liquid supply, sample positioning and the definition of the evaluation domain. The liquid was dispensed by a syringe pump (New Era NE-4000, precision of ±~2~\%) through a 0.7~mm Kapton tube at flow rates of 1–1.5 ml/h. The Kapton tube was vertically aligned relative to the sample carrier using an additively manufactured holder. Sample positioning relative to the X-ray beam was achieved using a micromechanical stage with a fitted pocket, while residual tilts or positional offsets below ±~50~µm could influence apparent absorption and contrast. To account for spatial heterogeneity, four cropped regions of interest were evaluated per image sequence and the resulting penetration constant~$C$ was averaged. The standard deviation between crops ranged from 0.251 to 6.959~$s^{-0.5}$, corresponding to relative variations of 2–16~\%. This variability quantifies the local stochastic fluctuations within each dataset. While no formal uncertainty propagation was performed, the combined influence of liquid-supply variability and alignment inaccuracy is assumed to remain within the observed experimental scatter. The reliability of the transport index~$T_z(\Theta)$ is therefore primarily limited by intrinsic material heterogeneity rather than by instrumental error.

\newpage
\section{Justification of the linear intensity normalization}
\label{Appendix:Linearization}
For each sample, $I_s$ is defined as the minimum greyscale intensity observed over the complete experiment, i.e.\ the most-wetted state reached across all applied droplets (lowest transmission). In Beer--Lambert form, the measured intensity can be written as

\begin{equation}
I(t)=I_0\,e^{(-\mu_0 L)}\,e^{(-\Delta\mu\,l(t))}
\end{equation}

where $e^{(-\mu_0 L)}$ groups approximately time-invariant contributions (solid matrix and setup), and $l(t)$ denotes the effective KI-solution path length along the beam. Here, $\Delta\mu$ denotes the incremental attenuation when pore space contains KI-solution instead of air, i.e.\ $\Delta\mu=\mu_{\mathrm{solution}}-\mu_{\mathrm{air}}\approx\mu_{\mathrm{solution}}$, since $\mu_{\mathrm{air}}$ is negligible at 16.6~keV over mm-scale path lengths. Using the effective attenuation length of the 30~wt\% KI solution at 16.6~keV, $\ell_{\mathrm{solution}}\approx 1.01~\mathrm{mm}$, we obtain

\begin{equation}
\mu_{\mathrm{solution}}\approx \frac{1}{\ell_{\mathrm{solution}}}\approx 0.99~\mathrm{mm^{-1}}
\end{equation}

and therefore $\Delta\mu\approx 0.99~\mathrm{mm^{-1}}$. Using the saturated reference

\begin{equation}
I_s=I_0\,e^{(-\mu_0 L)}\,e^{(-\Delta\mu\,l_s)}
\end{equation}

yields

\begin{equation}
I(t)=I_s\,e^{[-\Delta\mu\,(l(t)-l_s)]}
\end{equation}

The overall experiment spans conditions from nearly dry to highly wetted, which is reflected in the full dry-to-saturated intensity ratios $I_d/I_s\approx 54$ (Ref), $37$ (High-NPI) and $7.8$ (Low-NPI), corresponding to $\ln(I_d/I_s)\approx 4.0$, $3.6$ and $2.1$, respectively. However, the redistribution analysis is performed on intensity variations relative to the fixed sample-specific reference~$I_s$ over a short post-burst time window ($200$ frames, i.e.\ $\approx$4--5~s). Over this fitting interval, the magnitude of the exponential term can be quantified directly from the measured mean crop intensities at the burst frame $I_{\mathrm{burst}}$ and after $200$ frames $I_{200}$ as

\begin{equation}
\Delta\mu\,\Delta l=\ln\!\left(\frac{I_{200}}{I_{\mathrm{burst}}}\right)
\end{equation}

where $\Delta l=l(t)-l_s$. Using the data for one crop $j_c$, we obtain $\ln(I_{200}/I_{\mathrm{burst}})\approx 1.17$-$1.53$ for High-NPI droplets, $\approx 1.16$-$1.41$ for Low-NPI droplets 2-5 and $\approx 1.77$-$2.39$ for sample Ref. Consequently, only moderate variations in in $|\Delta\mu\,\Delta l|$ are present during the process of liquid redistribution. Therefore, a first-order approximation allows for linear mapping between $I_d$ and $I_s$ as implemented in Eq.~\ref{Eq:IntensitySaturation}, while maintaining a robust monotonic relation between intensity and wetting

\begin{equation}
e^{(-\Delta\mu\,\Delta l)} \approx 1-\Delta\mu\,\Delta l
\end{equation}

\newpage

\section{Determined values for C for different crops 1-4 for samples~Ref,~High-NPI~and~Low-NPI}
\label{Sec:App1}
\begin{table}[htbp]
\caption{Determined values for $C$ for different crops 1--4 for samples~Ref,~High-NPI and~Low-NPI. Mean and standard deviation (Std.) are calculated per droplet.}
\label{tab:C-Fits}
\begin{tabular}{@{}lllllll@{}}
\toprule
\multicolumn{7}{c}{Sample Ref} \\ \midrule
\multirow{3}{*}{Droplet} 
  & \multicolumn{4}{c}{C} 
  & \multicolumn{2}{c}{} \\ \cmidrule(lr){2-7}
 & Crop 1 & Crop 2 & Crop 3 & Crop 4 & Mean & Std. \\
 & (30,200,5,1000) & (40,200,5,605) & (60,200,5,850) & (25,190,20,1000) & & \\ \midrule
1  & 13.274 & 12.965 & 12.566 & 12.953 & 12.940 & 0.251 \\
2  & 13.218 & 10.490 & 12.182 & 11.744 & 11.909 & 0.978 \\
3  & 12.821 & 11.248 & 11.288 & 11.686 & 11.761 & 0.636 \\
4  & 13.828 & 13.641 & 12.069 & 12.884 & 13.106 & 0.695 \\
5  & 17.339 & 20.630 & 15.588 & 15.202 & 17.190 & 2.143 \\
6  & 19.631 & 24.776 & 17.558 & 17.510 & 19.869 & 2.960 \\
7  & 27.628 & 33.087 & 23.898 & 25.923 & 27.634 & 3.414 \\
8  & 28.161 & 29.439 & 23.839 & 27.861 & 27.325 & 2.098 \\
9  & 38.761 & 36.312 & 32.416 & 40.370 & 36.965 & 3.000 \\
10 & 38.349 & 37.030 & 32.766 & 41.076 & 37.305 & 3.000 \\
11 & 51.586 & 46.535 & 44.198 & 57.000 & 49.830 & 4.460 \\
12 & 46.770 & 42.334 & 40.611 & 52.135 & 45.463 & 4.460 \\
13 & 48.700 & 42.066 & 41.389 & 55.257 & 46.853 & 5.631 \\
14 & 59.089 & 51.127 & 50.487 & 67.445 & 57.037 & 6.899 \\
15 & 59.289 & 48.398 & 47.874 & 63.999 & 54.890 & 6.959 \\ \midrule

\multicolumn{7}{c}{Sample High-NPI} \\ \midrule
\multirow{3}{*}{Droplet} 
  & \multicolumn{4}{c}{C} 
  & \multicolumn{2}{c}{} \\ \cmidrule(lr){2-7}
 & Crop 1 & Crop 2 & Crop 3 & Crop 4 & Mean & Std. \\
 & (10,360,10,170) & (10,345,0,150) & (20,350,20,150) & (0,350,0,150) & & \\ \midrule
1 & 12.167 & 14.569 & 13.398 & 14.348 & 13.621 & 0.948 \\
2 & 24.141 & 27.068 & 19.742 & 26.808 & 24.440 & 2.944 \\
3 & 23.950 & 29.977 & 26.225 & 26.833 & 26.746 & 2.153 \\
4 & 25.566 & 35.223 & 31.095 & 35.118 & 31.751 & 3.939 \\
5 & 21.617 & 36.234 & 32.076 & 36.265 & 31.548 & 5.982 \\ \midrule

\multicolumn{7}{c}{Sample Low-NPI} \\ \midrule
\multirow{3}{*}{Droplet} 
  & \multicolumn{4}{c}{C} 
  & \multicolumn{2}{c}{} \\ \cmidrule(lr){2-7}
 & Crop 1 & Crop 2 & Crop 3 & Crop 4 & Mean & Std. \\
 & (10,360,5,130) & (10,320,5,155) & (20,310,10,150) & (15,300,20,145) & & \\ \midrule
1 & 5.706  & 6.740  & 6.292  & 5.978  & 6.179 & 0.385 \\
2 & 9.069  & 13.020 & 11.835 & 10.672 & 11.149 & 1.460 \\
3 & 10.451 & 16.482 & 14.675 & 13.054 & 13.666 & 2.217 \\
4 & 9.170  & 17.887 & 15.944 & 13.463 & 14.116 & 3.258 \\
5 & 26.401 & 33.485 & 30.325 & 29.633 & 29.961 & 2.520 \\ \bottomrule
\end{tabular}
\end{table}

\newpage

\section{Fitted values for C for sample~Ref}
\label{Sec:App2}
\begin{figure}[htbp]
\centerline{\includegraphics[width=.9\textwidth]{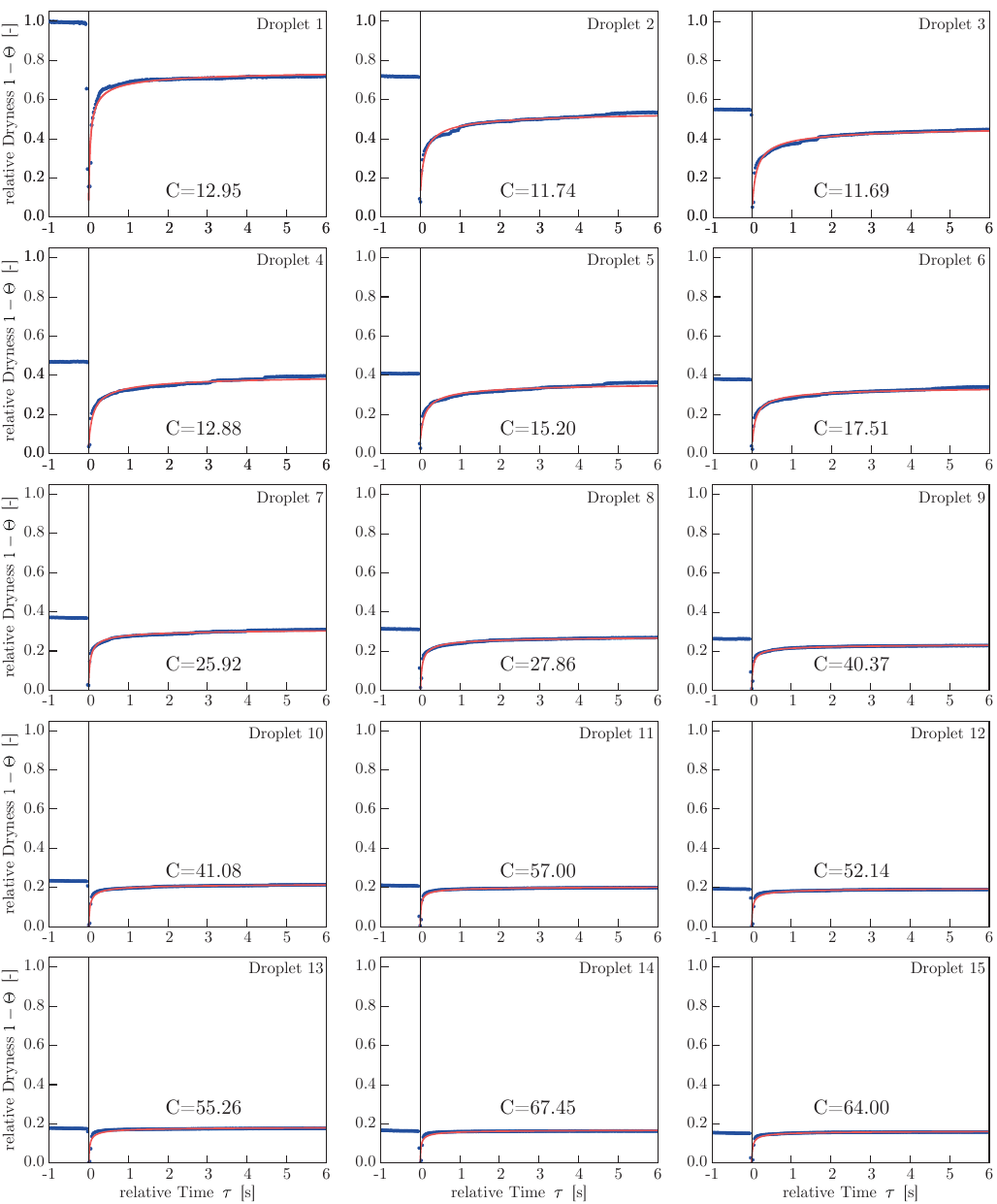}}
\caption{Results for the relative dryness $1-\Theta$ at sample~Ref during the subsequent release of 15 droplets at crop 1=(25,190,20,1000). For each droplet, the penetration constant~$C$ is determined by fitting the function in Eq.~\ref{Eqn:FitFunction} to the measured data. Measured data is depicted as blue points, whereas the fitted function is depicted as a red line.}
\label{Fig:Res_A_Overview}
\end{figure}

\newpage

\end{appendices}

\bibliography{sn-bibliography_cleaned}

\end{document}